\documentclass{iopart}
\usepackage{graphicx}
\usepackage{iopams}

\begin{document}
\title[Graphene NEM resonators for detection of modulated THz radiation]{Graphene nanoelectromechanical resonators for detection of modulated terahertz radiation}
\author{D.~Svintsov$^{1,2,3}$, V.G.~Leiman$^{2}$, V.~Ryzhii$^{3,4,5}$, T.~Otsuji$^{3}$ and M.S.~Shur$^6$}
\address{$^1$Institute of Physics and Technology, Russian Academy of Sciences, Moscow 117218, Russia}
\address{$^2$Moscow Institute of Physics and Technology (State University), Dolgoprudny 141700, Russia}
\address{$^3$Research Institute of Electrical Communication, Tohoku University, Sendai 980-8577,  Japan}
\address{$^4$Center for Photonics and Infrared Engineering, Bauman Moscow State Technical University, Moscow 105005, Russia}
\address{$^5$Institute of Ultra High Frequency Semiconductor Electronics, Russian Academy of Sciences, Moscow 111005, Russia}
\address{$^6$Center for Integrated Electronics and Department of Electrical, Computer and Systems Engineering, Rensselaer Polytechnic Institute, Troy, New York 12180, United States}
\ead{svintcov.da@mipt.ru}

\begin{abstract}
We propose and analyze the detector of modulated terahertz (THz) radiation based on the graphene field-effect transistor with mechanically floating gate made of graphene as well. The THz component of incoming radiation induces resonant excitation of plasma oscillations in graphene layers (GLs). The rectified component of the ponderomotive force between GLs invokes resonant mechanical swinging of top GL, resulting in the drain current oscillations. To estimate the device responsivity, we solve the hydrodynamic equations for the electrons and holes in graphene governing the plasma-wave response, and the equation describing the graphene membrane oscillations. The combined plasma-mechanical resonance raises the current amplitude by up to four orders of magnitude. The use of graphene as a material for the elastic gate and conductive channel allows the voltage tuning of both resonant frequencies in a wide range.
\end{abstract}

\submitto{\JPD}

\maketitle

\section{Introduction}
The resonant detection of radio signals using electromechanical systems demonstrated a long time ago~\cite{Nathanson} recently attracted a new wave of interest due to the advances in fabrication of nanoelectromechanical systems (NEMS) based on metal and semiconductor materials~\cite{Mems-review}, and, more lately, carbon-based structures~\cite{Feng-Graphene-Oscillator,Tunable-nanotube-oscillator}. Graphene, a two-dimensional allotrope of carbon, demonstrates unique mechanical properties, uppermost high elastic stiffness of $340$ N/m and ability to sustain a large mechanical stress (up to 40 N/m~\cite{Measurement-of-eleastic-properties}). The graphene-based NEMS oscillators exhibited resonant frequencies up to 260 MHz~\cite{Graphene-260-MHz-resonator} and are predicted to operate at frequencies up to tens of GHz~\cite{Chen-review}. Their tuning can be conveniently performed by changing the gate voltage~\cite{Tuning-mechanical-frequency}.

Graphene and other carbon materials also possess unique electronic properties. In the first place, it is high electron mobility~\cite{Bolotin} that allows ultrafast (up to THz) operation of graphene-based devices, including field-effect transistors (FETs)~\cite{THz-graphene-transistors}, optical modulators~\cite{Graphene-modulator,Plasma-resonances-in-modulator}, and detectors of radiation~\cite{THz-detection-graphene-FET}. High mobility also facilitates the resonant plasma wave excitation in those structures. For micron-length graphene-resonators, the eigenfrequency of plasma oscillations lies in the THz range~\cite{Ryzhii-plasmons-2}. The excitation of plasma waves can significantly increase the efficiency of THz detection using transistor-like structures and provide highly selective (resonant) response~\cite{Dyakonov-Shur-Detection}.

In this paper, we propose the resonant detector of modulated THz radiation which exploits {\textit{both}} unique mechanical and electronic properties of graphene. The necessity for resonant transduction of modulated THz signals can appear in future telecommunication systems, where the THz carrier frequencies are expected to allow for higher transmission rate.
\begin{figure}[ht]
\center{\includegraphics[width=1.0\linewidth]{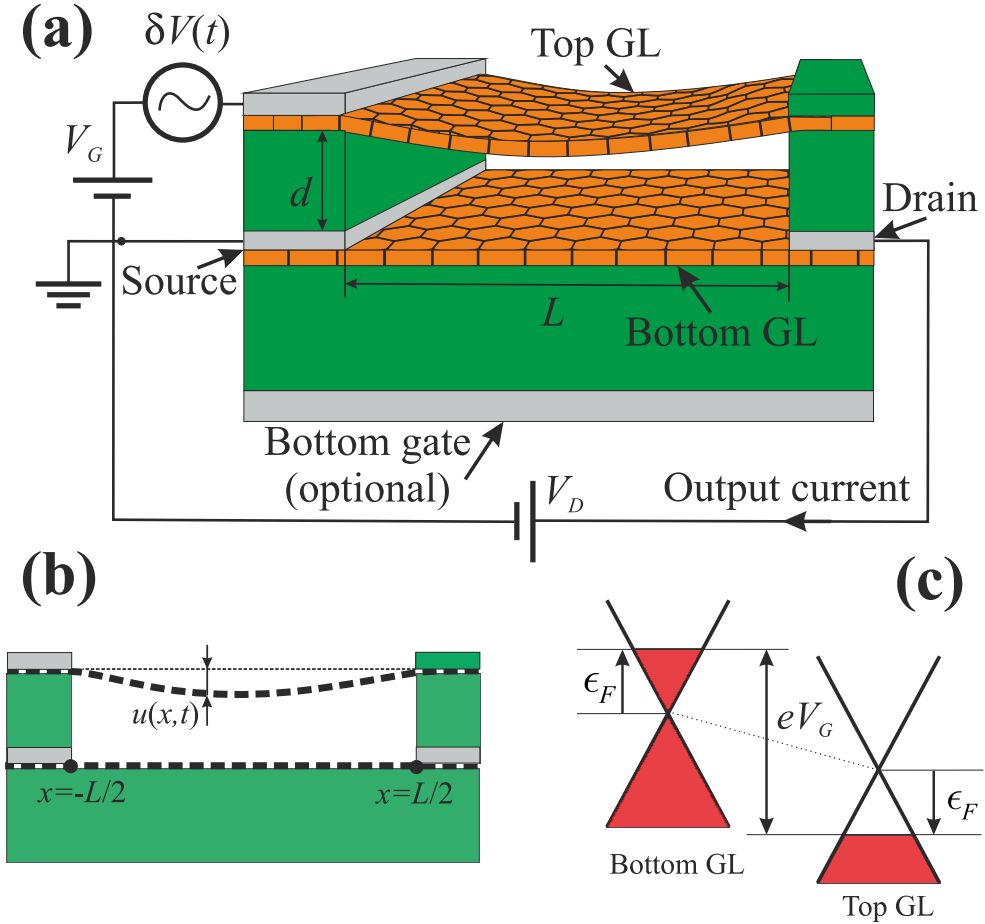} }
\caption{(a) Sketch of the proposed detector structure (b) Schematic view of the deflection $u(x,t)$ of suspended top GL (c) Band diagram of double-GL structure under applied voltage $V_G$ (filled areas are occupied by electrons)}
\label{F1}
\end{figure}

The proposed device structure represents a graphene FET with a mechanically floating gate made of graphene as well [figure~\ref{F1} (a)]. The carrier signal is impinging on the structure or is delivered via a waveguide. The amplitude-modulated signal $\delta V(t)$ with the modulation frequency $\omega_m$ is applied between source and gate contacts. The carrier frequency $\omega$ is in the THz range being close to the Eigen frequency $\Omega$ of the plasma oscillations. The modulation frequency $\omega_m$ is in the GHz range, which is close to the frequency $\Omega_m$ of the top gate mechanical oscillations.

The carrier frequency excites plasma oscillations, which result in increasing the electric field strength between the top and bottom GLs. This, in turn, leads to a large ponderomotive force between them. The force spectrum contains the rectified component oscillating at the modulation frequency, which invokes mechanical oscillations of the graphene gate [figure~\ref{F1}~(b)]. The output signal of the detector is the ac source-drain current varying due to the changing gate-to-channel capacitance.

The THz demodulators exploiting combined plasma-mechanical resonance were first proposed in Refs.~\cite{Ryzhii2007-HEMT-detector,Nanotube-guitar,Nanotube-combined-resonance}. Those devices incorporated a conductive mechanically floating cantilever~\cite{Ryzhii2007-HEMT-detector}, nanowire or a nanotube~\cite{Nanotube-guitar} suspended over the channel of high-electron-mobility transistor. The structures of two aligned nanotubes were also studied~\cite{Nanotube-combined-resonance}. The use of graphene in such kind of a device allows one to attain higher electron mobility, and thus a higher responsivity. Large breaking strain of graphene~\cite{Measurement-of-eleastic-properties,Chen-review} enables the gate rigidity tuning in a wide range, which is hardly possible for metallic cantilevers.

To estimate the plasma-wave response of the device, we apply the hydrodynamic equations for massless electrons and holes in graphene~\cite{Our-hydrodynamic}. The mechanical vibrations are modelled using the elasticity theory equations for graphene membranes~\cite{Elasticity-Theory,Continuum-elastic}. We show that the resonant responsivity of modulated radiation detection is proportional to $Q_m Q_p^2$ where $Q_m$ and $Q_p$ are the quality factors of mechanical and plasma resonators, respectively. We also show that the resonant amplitude of source-drain current oscillations is proportional to the third power of the electron mobility. Thus, the device responsivity appears to be large, reaching $\sim 10^3$ A/W in the resonant case. The gate voltage can effectively tune both the plasma and mechanical resonant frequencies. The optional bottom gate below the whole structure would allow for independent control of those frequencies.  The device could be also used as an element of a mixer or a heterodyne detector if two THz frequencies are fed in and the difference is in resonance with the mechanical Eigen frequency. It can also operate as a detector of non-modulated THz radiation, with the responsivity by a factor of $Q_m$ smaller than that in the case of the modulated radiation.

The paper is devoted to the analytical model describing the device output current and responsivity. Section II deals with the plasma-wave response of the structure. In Section III, the mechanical oscillations of the suspended graphene gate are considered. In Section IV, we estimate the output current of the device and its responsivity, and discuss possible generalizations of the considered structures. Section V contains the main conclusions.

\section{Plasma-wave response}

The proposed device consists of two GLs with the top layer being suspended over the bottom layer [figure~\ref{F1} (a)]. The length and width of he layers are $L$ and $W \gg L$, respectively, the distance between the GLs is $d$. The gate voltage including a constant DC bias $V_G$ and the amplitude-modulated signal 
\begin{equation}
\label{Signal}
\delta V(t) = \delta V_m \cos(\omega t) [1 + m \cos(\omega_m t)]
\end{equation}
is applied between left-side contacts to GLs ($m$ is the modulation depth). A bias voltage $V_D$ (drain voltage) is applied across the bottom GL, allowing for dc current flow to be modulated by the incoming signal. The right contact to the top GL is eclectically isolated (or it could be connected to the top left contact).

Application of the DC gate voltage $V_G$ leads to the accumulation of electrons and holes in the opposite layers, as shown in figure~\ref{F1}~(c). In the absence of a built-in voltage, the Fermi energies $\epsilon_F$ of the electrons and holes are equal in modulus and opposite in sign. The electron and hole two-dimensional sheet densities $\Sigma$ are related to $V_G$ via a local capacitance relation
\begin{equation}
\label{Steady_electrostatics}
C V_G= e \Sigma,
\end{equation}
where $C = (C_g C_q/2)/(C_g + C_q/2)$ is the effective specific capacitance corresponding to the series connection of geometric capacitance $C_g$ and quantum capacitances. The geometric capacitance per unit area is $C_g = \varepsilon_0/d$. The quantum capacitance, defined by $C_q = e^2 \partial \Sigma/\partial \epsilon_F$~\cite{Quantum-capacitance} accounts for the dependence of the Fermi energy on the electric field strength between the GLs [figure~\ref{F1} (c)]. For large distances between GLs ($d\gtrsim 20$ nm at room temperature) or for high carrier densities $C_q$ is much larger than the geometric capacitance. At such conditions, $C \approx C_g$, which will be assumed in the following.
  
Consider the plasma wave response of the transistor structure in figure~\ref{F1}~(a) to the application of a small harmonic signal $\delta V e^{-i \omega t}$ at the left end of the top GL. The resulting voltage difference between the top and bottom layers,  $\delta \varphi_+ - \delta \varphi_-$, is related to the perturbation of the charge density $e \delta \Sigma$ via
\begin{equation}
\label{Electrostatic}
C\left( \delta {\varphi_-}-\delta {\varphi_+} \right) = e \delta \Sigma.
\end{equation}
Combining the Ohm's law with the continuity equations for the top and bottom GLs, we can relate the density perturbation $\delta\Sigma$ to the perturbations of top and bottom layer potentials $\delta\varphi_\pm$: 
\begin{eqnarray}
\label{Continuity-1}
i\omega e \delta \Sigma = \sigma_+ \frac{\partial^2\delta \varphi_+}{\partial x^2},\\
\label{Continuity-2}
i\omega e \delta \Sigma = - \sigma_- \frac{\partial^2\delta \varphi_-}{\partial x^2},
\end{eqnarray}
where $\sigma_+$ and $\sigma_-$ are the sheet conductivities of top and bottom GLs. The spatial variation of conductivity is assumed to be weak, which is valid at small drain voltage $V_D < V_G$ (i.e. in the linear mode of FET). Because of the equal carrier densities in the GLs and electron-hole symmetry $\sigma_+ = \sigma_- = \sigma$. A slight deviation from this identity due to a stronger disorder in the bottom GL is possible. This issue will be briefly discussed later.

Combining (\ref{Electrostatic}), (\ref{Continuity-1}), and (\ref{Continuity-2}), we arrive at the equations governing the voltage distribution along the GLs
\begin{eqnarray}
\label{Dynamic_Eqs-1}
\frac{\partial^2\delta \varphi_+}{\partial x^2} = \frac{i\omega C }{\sigma} \left( \delta {\varphi_-}-\delta {\varphi_+} \right),\\
\label{Dynamic_Eqs-2}
\frac{\partial^2\delta \varphi_-}{\partial x^2} = \frac{i\omega C }{\sigma} \left( \delta {\varphi_+}-\delta {\varphi_-} \right).
\end{eqnarray}
The boundary conditions imply constant ac voltage $\delta V$ at the left edge of the top GL, zero ac voltage at both ends of bottom GL, and zero current at the isolated edge of top GL:
\begin{eqnarray}
{\left. \delta {\varphi_+} \right|}_{x=-L/2} = \delta V,\nonumber \\
\label{BC}
{\left. \delta \varphi_- \right|}_{x=\pm L/2} = 0,\qquad \left. (\partial\delta \varphi_+/\partial x) \right|_{x=L/2} = 0.
\end{eqnarray}
The quantity $i \omega C / \sigma$ in Eqs.~(\ref{Dynamic_Eqs-1}--\ref{Dynamic_Eqs-2}) has the dimensionality of the wave vector squared, we denote it by $\gamma_{\omega}^2/2$. Using the Drude-like expression for the graphene conductivity $\sigma = \left[e^2 \epsilon_F /(\pi \hbar^2 \nu)\right] \left[ 1 + i \omega/\nu \right]^{-1}$, where $\nu$ is the carrier collision frequency~\cite{Our-hydrodynamic}, we obtain the frequency dependence of $\gamma_{\omega}$:
\begin{equation}
\gamma_{\omega} = \frac{\sqrt{2 \omega (\omega + i \nu)}}{s},
\end{equation}
where $s = \left[e^2 \epsilon_F / (\pi \hbar^2 C)\right]^{1/2}$ is the velocity of plasma waves in the gated graphene~\cite{Ryzhii-plasmons-2,Our-hydrodynamic}. Equations (\ref{Dynamic_Eqs-1}--\ref{Dynamic_Eqs-2}) are then rewritten as
\begin{eqnarray}
\label{Transport-1}
\frac{\partial^2\delta\varphi_+}{\partial x^2} + \frac{\gamma^2_\omega}{2} \left( \delta \varphi_+ - \delta \varphi_- \right),\\
\label{Transport-2}
\frac{\partial^2\delta\varphi_-}{\partial x^2} + \frac{\gamma^2_\omega}{2} \left( \delta \varphi_- - \delta \varphi_+ \right).
\end{eqnarray}

\begin{figure}[ht]
\center{\includegraphics[width=1.0\linewidth]{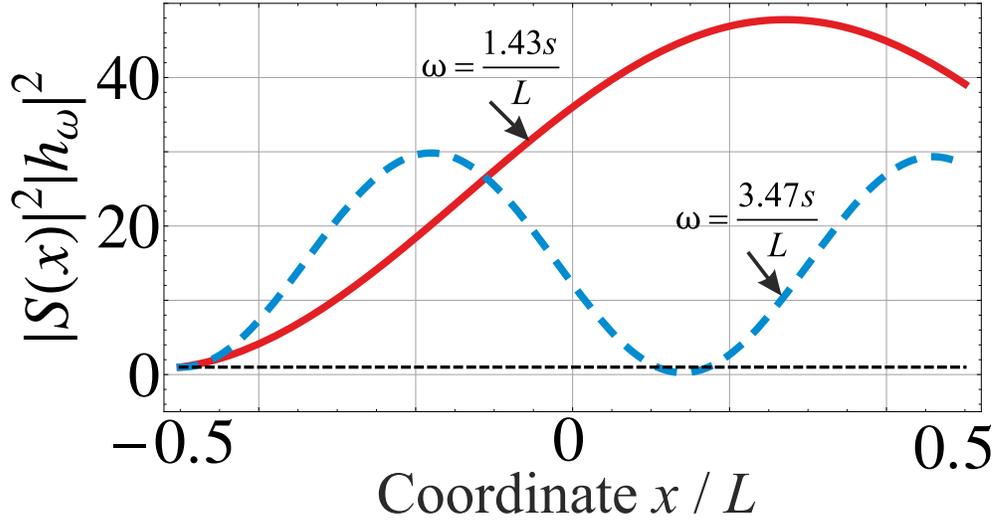} }
\caption{Spatial profile of ponderomotive force $|h_\omega S(x)|^2$ vs. coordinate $x/L$. Red and blue lines correspond to the first and second resonances, respectively. Black dashed line indicates non-resonant ($\omega = 0$) constant force profile}
\label{F2}
\end{figure}

Solving~(\ref{Transport-1}--\ref{Transport-2}) with boundary conditions (\ref{BC}), we find the voltage difference between GLs
\begin{equation}
\label{Voltage-distribution}
\delta \varphi_+ - \delta \varphi_- = \delta V h_\omega S(x),
\end{equation}
where $h_\omega$ is the dimensionless plasma resonant factor
\begin{equation}
h_\omega = \left[\cos \left( \gamma_\omega L \right) + \frac{\sin ( \gamma_\omega L )}{\gamma_\omega L }  \right]^{-1},
\end{equation} 
and $S(x)$ is the dimensionless function describing the spatial AC voltage distribution
\begin{equation}
S(x)  =  \cos\left(\frac{\gamma_\omega L}{2} -\gamma_\omega x\right) + \cos(\gamma_\omega x) \frac{\sin(\gamma_\omega L/2)}{\gamma_\omega L/2} .
\end{equation}
The spatial profiles of ponderomotive force proportional to $|h_\omega|^2 |S(x)|^2$ are shown in figure~\ref{F2} for the two first plasma resonances. 

The resonant frequencies in the limit of weak damping $\nu \ll \omega$ are found from the solution of the transcendental equation $x \cos x + \sin x = 0$, where $x = \gamma_\omega L$. The first root of this equation, $x_0 \approx 2.03$, corresponds to the plasma resonant frequency $\Omega = 1.43 s/L$. In the vicinity of this resonance, $|h_\omega|^2$ is described by the Lorentz function
\begin{equation}
|h_\omega|^2\approx \frac{0.54 Q_p^2}{1 + 4 Q_p^2 (\omega/\Omega - 1)^2},
\end{equation}
where we have introduced the quality factor of the plasma oscillations $Q_p = \Omega/\nu$. For a typical value of the plasma wave velocity $s\approx 4.5 \times 10^6$ m/s which corresponds to the gate voltage of 2 V across the distance $d=50$ nm, the resonant frequency $\Omega/2\pi \approx 1$ THz.

\begin{figure}[ht]
\center{\includegraphics[width=1.0\linewidth]{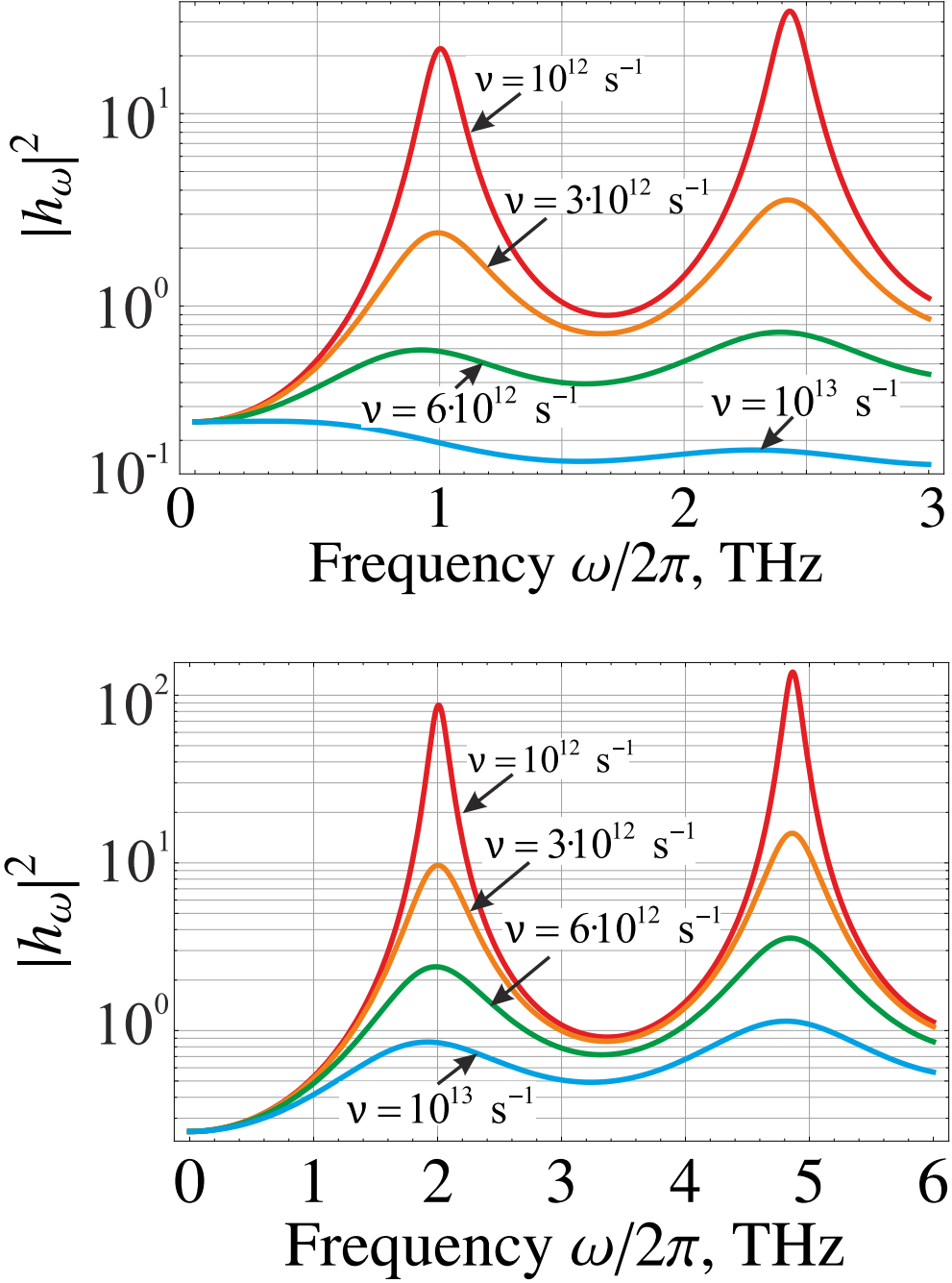}}
\caption{Plasma response function $|h_\omega|^2$ vs. carrier frequency $\omega$ at different values of collision frequency $\nu$. The structure length $L = 1$ $\mu$m (top) and $L = 500$ nm (bottom), the velocity of plasma waves $s=4.5 \times 10^6$ m/s.}
\label{F3}
\end{figure}

The magnitude and width of plasma resonance peaks are governed by the collision frequencies in the GLs. A careful analysis shows that if the collision frequencies $\nu_\pm$ in the top and bottom GLs are different, their average value $(\nu_+ + \nu_-)/2$ appears in the expression for $Q_p$. In a suspended graphene layer, the collision frequency is quite small being limited by acoustic phonon scattering~\cite{Bolotin}. Provided $\epsilon_F \gg T$, it can be estimated~\cite{Vasko-Ryzhii} as $\nu = \nu_0 (\epsilon_F/k_BT)$, where $\nu_0 \simeq 3.5 \times 10^{11}$ s$^{-1}$. The collision frequency in the bottom GL is larger due to scattering on substrate defects. The reported~\cite{Graphene-on-HBN} room-temperature electron mobility in graphene on boron nitride substrates is as high as $10^4$~cm$^2/$(V s), which is only an order of magnitude lower than the mobility in suspended samples. In our calculations, we use the collision frequency $\nu$ as a free parameter, keeping in mind that the values reaching $10^{12}$ s$^{-1}$ are possible. The plasma resonant curves in figure~\ref{F3} are plotted for the two values of the length of the structure, $L=1$~$\mu$m and $L=500$~nm, and for the collision frequencies varying from $10^{12}$ s$^{-1}$ to $10^{13}$ s$^{-1}$. Despite a rather low Q-factor ($Q_p \approx 6$ for $\nu = 10^{12}$~s$^{-1}$ and $L=1$~$\mu$m), the resonant detection is still pronounced.

Having obtained the plasma-wave response~(\ref{Voltage-distribution}), we find the average ponderomotive force $\overline{f(x,t)}$ acting between the GLs (the averaging is performed over time $2\pi/\omega \ll \tau \ll 2\pi/\omega_m$):
\begin{eqnarray}
\label{Force}
\fl \overline{f(x,t)} = \frac{CV_G^2}{2d} + \frac{CV_G^2}{d^2} \delta u(x,t) +\nonumber \\
|h_\omega|^2|S(x)|^2\frac{C\delta V_m^2}{4 d} [1 + m \cos(\omega_m t)]^2.
\end{eqnarray}
The first term in equation (\ref{Force}) is the attraction force due to the constant gate voltage $V_G$. The second term is due to the varying distance $\delta u(x,t)$ between the electrodes at a fixed voltage. The third term is due to the rectification of the amplitude-modulated signal. It contains three harmonics: zero-frequency, which can be used for detection of non-modulated signals; the harmonic with the modulation frequency $\omega_m$ which leads to the forced oscillations of the top GL and invokes a mechanical resonance; and the double-frequency harmonic $2\omega_m$.

\section{Mechanical response}
The deflection, $u(x,t)$, of the suspended GL is found from the solution of the elasticity equations for graphene. According to the theory developed in Ref.~\cite{Continuum-elastic}, the bending energy of graphene is much less than its stretching energy. Thus, the time-dependent deformation of the GL is governed by the wave equation of the membrane oscillations. Presenting the deflection $u(x,t)$ of the top GL as $u_0(x) + \delta u(x,t)$, where $\delta u(x,t)$ oscillates with the modulation frequency, $\delta u (x,t) = \delta u_m(x) e^{-i\omega_m t}$, we write this equation as
\begin{eqnarray}
\label{Wave-fourier}
\fl \rho \omega_m \left( \omega_m  - i \nu_m \right) \delta u_m +  T \frac{\partial^2 \delta u_m}{\partial x^2} = \nonumber \\
-\frac{C V_G^2}{d^2} \delta u_m - \frac{m}{2} \frac{C \delta V_m^2}{d} |h_\omega|^2 |S(x)|^2.
\end{eqnarray}
Here, $\rho = 7 \times 10^{-7}$ kg/m$^2$ is the mass density of graphene, the frequency $\nu_m$ phenomenologically accounts for mechanical damping, and $T$ is the elastic force density. The latter is proportional to the tensile strain $\delta_x$ of the graphene layer:
\begin{equation}
T = E h \delta_x,
\end{equation}
where $E h = 340$ N/m is the two-dimensional elastic stiffness.

The first force term in the right-hand side of equation~(\ref{Wave-fourier}) is known to pull the resonant frequencies to lower values~\cite{Chen-review,Tuning-mechanical-modes}. We denote the corresponding frequency shift as $\omega_{sh} = (V_G/d)(C/ \rho)^{1/2}$ (for $V_G = 2$~V and $d=50$~nm $\omega_{sh}/2\pi \approx 400$ MHz). Introducing the modified mechanical frequency $\tilde{\omega}_m^2 = \omega_m (\omega_m - i\nu_m) + \omega^2_{sh}$ and the velocity of transverse sound $c_s = \sqrt{T/\rho}$, we present equation~(\ref{Wave-fourier}) in a concise form
\begin{eqnarray}
\label{Wave-fourier-2}
\tilde{\omega}_m^2 \delta u_m +  c_s^2 \frac{\partial^2 \delta u_m}{\partial x^2} = - \frac{m}{2} \frac{C \delta V^2}{d} |h_\omega|^2 |S(x)|^2.
\end{eqnarray}
The solution of equation~(\ref{Wave-fourier-2}) with zero boundary conditions (clamped edges) is straightforward, moreover, for the estimate of change in gate-to-channel capacitance we need to know only the deflection averaged over $x$-coordinate $\langle \delta u_m (x) \rangle$. The latter can be presented as
\begin{equation}
\label{Average-deflection}
\langle \delta u_m (x) \rangle = \frac{m}{4} \frac{C \delta V^2}{\rho d |\tilde \omega_m^2|} |h_\omega|^2 |H_{\omega, \omega_m}|.
\end{equation}
Here we have introduced the mechanical resonant factor $|H_{\omega, \omega_m}|$. In the limit of weakly damped plasma oscillations, $\nu \ll \Omega$,  it is given by the following expression
\begin{eqnarray}
\label{Mechanical-resonance}
\fl H_{\omega ,\omega_m } =  \frac{\tan \gamma_m L}{\gamma_m L} \left[ \alpha_- \frac{\cos \gamma'_\omega L} {1 - \left( \gamma'_\omega / \gamma_m \right)^2} + \alpha_+ \right] - \nonumber \\
 \left[ \alpha_- \frac{\sin ( \gamma'_\omega L )/ ( \gamma'_\omega L )}{1-  \left( \gamma'_\omega /\gamma_m \right)^2} + \alpha_+ \right] ,
\end{eqnarray}
where $\gamma_m = \tilde{\omega}_m/(2 c_s)$, $\gamma'_\omega$ is the real part of $\gamma_\omega$, and its imaginary part is assumed to be small, $\alpha_\pm$ are non-resonant factors,
\begin{equation}
\alpha_\pm = \left[\cos(\gamma'_\omega L/2) + \frac{\sin(\gamma'_\omega L/2)}{\gamma'_\omega L/2} \right] \pm \sin^2(\gamma'_\omega L/2).
\end{equation}

\begin{figure}[ht]
\center{\includegraphics[width=1.0\linewidth]{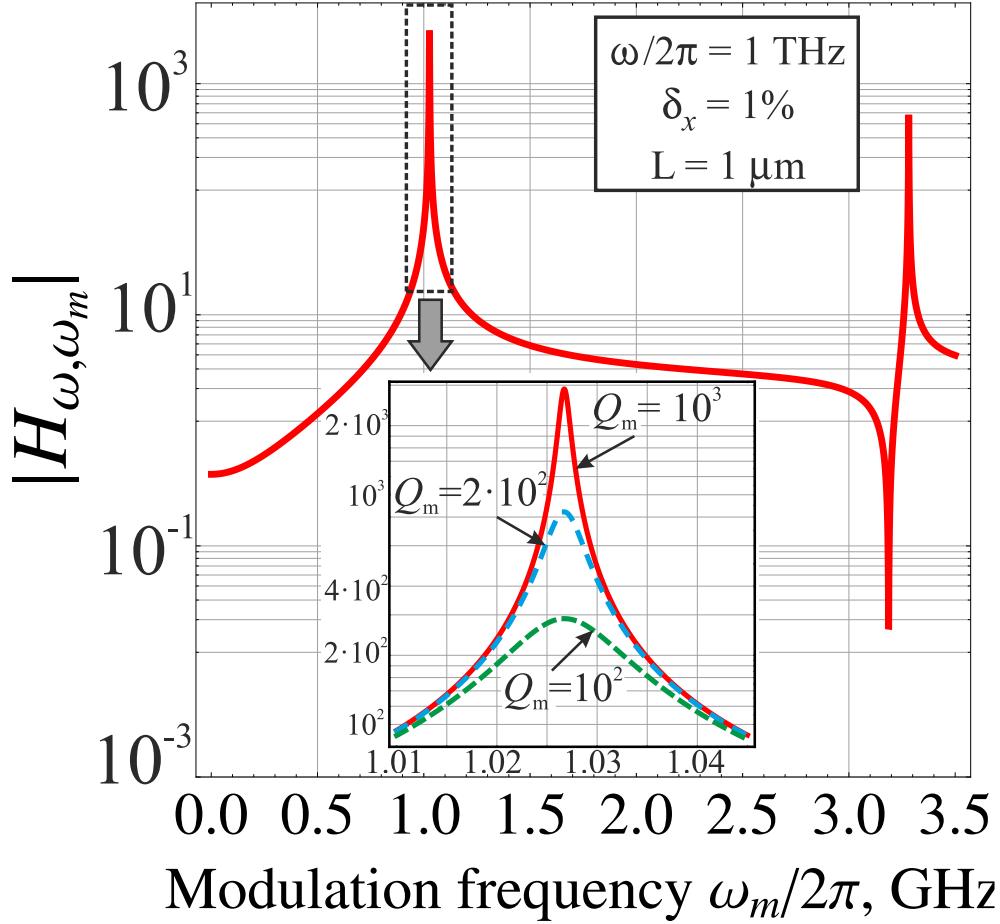} }
\caption{Mechanical response function $|H_{\omega,\omega_m}|$ vs. modulation frequency $\omega_m$. Structure length $L = 1$ $\mu$m, tensile strain of top GL is $\delta_x = 0.01$, quality factor $Q_m = 10^3$. Inset: response functions in the vicinity of resonance for $Q$-factors ranging from $10^2$ to $10^3$}
\label{F4}
\end{figure}

The dependence of mechanical resonant factor $H_{\omega ,\omega_m }$ provides a complicated pattern of resonances and anti-resonances appearing due to the spatial distribution of driving force (see figure~\ref{F4}). The principal resonant dependence is given by the term $\tan \left( \gamma_m L \right)$. In the vicinity of the first mechanical resonant frequency $\Omega_m = \pi c_s / L$ and plasma resonance, $|H_{\omega ,\omega_m }|$ is described by the Lorentzian function
\begin{equation}
|H_{\Omega ,\omega_m }| \approx \frac{1.34 Q_m}{\sqrt{1 + 4 Q_m^2\left(1 - \frac{\displaystyle{\sqrt{\omega_m^2 + \omega_{sh}^2}}}{\displaystyle{\Omega_m}}\right)^2}}.
\end{equation}
Here we have introduced the quality factor of mechanical resonator $Q_m = \Omega_m/\nu_m$. The reported~\cite{Chen-review} room-temperature quality factors for the graphene resonators tuned to frequencies of hundred MHz are of the order of $10^2$. In figure~\ref{F4} we present the calculated mechanical response function for the tensile strain $\delta_x = 1$ \%, $L = 1$~$\mu$m, and quality factors $Q_m$ ranging from $10^2$ to $10^3$. These values correspond to the resonant frequency $\Omega_m \approx 1$ GHz. Higher frequencies are attainable for the all-clamped graphene structures and for GLs strained by a sufficiently large gate voltage. 

\section{Results: output current, responsivity, and tuning}

Now we can estimate the amplitude of source-drain density $\delta {\cal J} (\omega_m)$ resulting from the varying gate-to-channel capacitance. Provided that the FET operates in the linear mode (the constant drain voltage $V_D$ is smaller than the gate voltage $V_G$), $\delta {\cal J} (\omega_m)$ is given by
\begin{equation}
\label{Drain-current-variation}
\delta {\cal J} (\omega_m) = \mu C V_G \frac{\langle \delta u_m (x) \rangle}{d} \frac{V_D}{L},
\end{equation}
where $\mu$ is the carrier mobility in the bottom GL. Using the obtained frequency dependence of the top GL deflection (\ref{Average-deflection}), we ultimately find
\begin{equation}
\label{Current_response}
\delta {\cal J} ( \omega_m ) = \frac{m}{4} \frac{\mu C V_G V_D}{L} \frac{C \delta V^2}{\rho |\tilde\omega_m|^2 d^2} |h_\omega|^2 |H_{\omega ,\omega_m }| .
\end{equation} 

\begin{figure}[ht]
\center{\includegraphics[width = 1.0\linewidth]{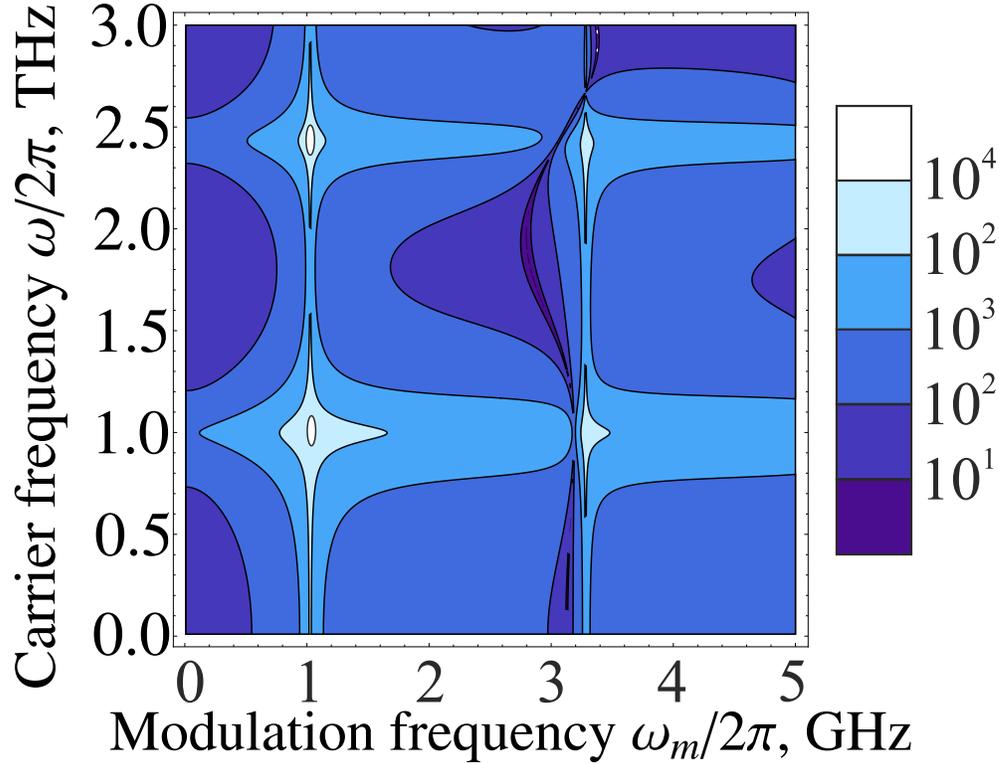} }
\caption{Dependence of combined plasma-mechanical response function $|h_\omega|^2 |H_{\omega\,\omega_m}|$ on carrier and modulation frequencies. The following parameters are used $\nu = 10^{12}$~s$^{-1}$, $\nu_m = 10^7$~s$^{-1}$ ($Q_m = 7 \times 10^2$), $L=1$~$\mu$m (log scale)}
\label{F5}
\end{figure}

The quantity $\mu C V_G V_D/L$ is the steady-state charge accumulated on a single GL per unit width $W$, divided by the electron drift time from the source to the drain. Under the conditions of the combined plasma and mechanical resonance ($\omega = \Omega\, , \omega_m = \Omega_m$), the detector current density is proportional to $Q_m Q_p^2$:
\begin{equation}
\label{Current_response_resonant}
\left.\delta {\cal J} (\Omega_m) \right|_{\omega=\Omega} = 0.18 m \frac{\mu C V_D V_G}{L} \frac{C \delta V^2} {\rho |\tilde\omega_m|^2 d^2} Q_p^2 Q_m .
\end{equation}

In figure~\ref{F5}, we plot the dependence of combined plasma-mechanical resonant factor $|h_\omega|^2 |H_{\omega\,\omega_m}|$ on the carrier and modulation frequencies. It exhibits sharp peaks in the vicinity of $\omega \approx \Omega$ and $\omega_m \approx \Omega_m$, with the resonant value exceeding $10^4$. It follows from equation~(\ref{Current_response_resonant}) that the current response is proportional to the third power of electron mobility (the second power comes from plasma quality factor squared and the first power comes from the drift time). Thus, the proposed device fully exploits the unique high-frequency electronic properties of graphene. The gate and channel in the proposed structure can be interchanged, so that the signal current flows in a suspended layer with enhanced electron mobility.

It should be noted that the proposed device can serve as a detector or non-modulated THz radiation. In this case, the rectified (zero-frequency) component of ponderomotive force would result in changing the DC source to drain current. The change in the DC current in the presence of a THz signal can be estimated as
\begin{equation}
\label{Current-response-non-modulated}
\left. \delta {\cal J} (\omega_m = 0) \right|_{\omega=\Omega} \approx  0.1 \frac{\mu C V_G V_D}{L} \frac{C \delta V^2} {\rho \Omega_m^2 d^2} Q_p^2, 
\end{equation}
which differs from Eq.~(\ref{Current_response_resonant}) by the factor of $Q_m$. 

Using Equations~(\ref{Current_response_resonant}) and (\ref{Current-response-non-modulated}), we estimate the detector responsivity $R = \delta {\cal I}/ \delta P$, where $\delta P$ is the power of incoming THz radiation received by the antenna, and $\delta {\cal I} = W \delta {\cal J}$ is the detector current. We use the typical parameters to attain the plasma and mechanical resonances at THz and GHz frequencies, respectively: $L = 1$ $\mu$m, $d=50$ nm, $V_G = 2$ V, $V_D = 1$ V, $Q_p \approx 6$, $Q_m \approx 10^3$, $W=10 L$, $m=1$. Relating $\delta P$ to $\delta V^2$ via 
\begin{equation}
\delta P = \frac{2 G c \varepsilon_0}{\pi} \delta V^2
\end{equation}
where $G$ is the antenna gain (for dipole antenna $G \approx 1.5$), we obtain the responsivity of $10^3$ A/W for the detection of the modulated radiation, and $0.6$ A/W for detection of non-modulated radiation. The responsivity might be greatly enhanced by increasing $V_D$ due to a large non-linearity in the FET saturation regime.

Both plasma and mechanical resonant frequencies $\Omega$ and $\Omega_m$ can be tuned by application of the constant top gate voltage $V_G$ since it controls the electron and hole densities in the GLs [Eq.~(\ref{Steady_electrostatics})] and their Fermi energies $\epsilon_F$. The plasma wave velocity is $s = \left[e^2 \epsilon_F / (\pi \hbar^2 C)\right]^{1/2}$. Thus, the plasma resonant frequency is a slowly increasing function of the gate voltage, $\Omega\propto V_G^{1/4}$. The dependence of the mechanical resonant frequency on $V_G$ is more complicated~\cite{Tuning-mechanical-frequency,Tuning-mechanical-modes}. On one hand, the application of the gate voltage increases $\omega_{sh}$ and pulls the resonant frequency to lower values. However, at low voltages another factor is more important. The tensile strain of GL $\delta_x$ includes the built-in strain $\delta_{x0}$ and voltage-induced strain $1/2 (\partial u_0(x)/\partial x)^2$, which is a growing function of $V_G$. 

Not going deep into mathematical details, we note that the addition of a back gate below the entire structure makes the independent tuning of mechanical and plasma resonances plausible. In this case, $\Omega_m$ will still depend only on top gate voltage, $V_G$, while the plasma wave velocity will be a function of both top and bottom gate voltages.

This device might also find applications as an extremely sensitive mass, gas~\cite{Graphene-Gas} or biosensor~\cite{Graphene-Bio}. The advantage in mass sensing compared to NEMs is a much larger area of a graphene gate compared to a nanowire or a nanotube. The sensing responsivity is also greatly enhanced in THz plasmonic devices~\cite{Graphene-Bio} because of a high plasma wave sensitivity to changes in the electric field distribution caused by the sensed medium.

Finally, we note that for the operation of the proposed THz detector, low-resistivity contacts to graphene layers are required. Recently, contact resistances of $500$ Ohm$\cdot\,\mu$m were reported for graphene/nickel contacts~\cite{Performance-killer}, while special treatment procedures can reduce this value down to $100$--$200$ Ohm$\cdot\,\mu$m~\cite{Low-contact-resistance-1,Low-contact-resistance-2}. These values correspond to the recharging frequency of the order of $10^{13}$ s$^{-1}$ for $L=1$ $\mu$m and $d=50$ nm, which allows the device operation in the THz range.

\section{Conclusions}
We have proposed and substantiated the operation of a resonant detector of the THz radiation modulated by GHz-signals. The device uses a graphene field-effect transistor with mechanically floating graphene gate. The THz component of incoming radiation invokes plasma resonance in the graphene layers, thus leading to a high ponderomotive force. The component of the ponderomotive force oscillating with the modulation frequency excites mechanical vibrations of the graphene gate. This leads to the change in the source-drain current. The resonant responsivity is proportional to $Q_p^2 Q_m$. For the structures that are $0.5...1$ $\mu$m long, the values $Q_p \approx 10$ and $Q_m \approx 10^3$ look feasible. Frequencies of both plasma and mechanical oscillations can be tuned by a constant gate voltage.

\section*{Acknowledgement}
The work at RIEC was supported by the Japan Society for Promotion of Science (JSPS Grant-in-Aid for Specially Promoting Research $\#$ 23000008), Japan. The work of D.S. was supported by the JSPS Postdoctoral Fellowship for Foreign Researchers (Short-term), Japan. The work of V.L. was supported by the Russian Foundation for Basic Research (grants \# 12-07-00710, 12-07-00592, and 13-07-00270). The work at RPI was supported by the US Army Cooperative Research Agreement (Program Manager Dr. Meredith Reed).

\end{document}